# Reduction of self-heating effects in GaN HEMT via h-BN passivation and lift-off transfer to diamond substrate: a simulation study

Fatima Z. Tijent[*,1,2,3], Mustapha Faqir[4], Paul L. Voss[2,3], Jean-Paul Salvestrini[2,3] and Abdallah Ougazzaden[2,3]

[*] *Corresponding author*

[1]*International University of Rabat, Aerospace and Automotive Engineering School and LERMA Lab, Rabat-Salé, 11100, Morocco*

[2]*CNRS, Georgia Tech – CNRS IRL 2958, 2 rue Marconi, 57070 Metz, France*

[3]*Georgia Institute of Technology, School of Electrical and Computer Engineering, Atlanta, GA 30332-0250, USA*

[4]*International University of Rabat, Aerospace and Automotive Engineering School, Rabat-Salé, 11100, Morocco*

**Abstract**

In this article, we investigate through numerical simulation the reduction of self-heating effects (SHEs) in GaN HEMT via the integration of hexagonal boron nitride (h-BN) as a passivation layer and as a release layer to transfer GaN HEMT to diamond substrate. The obtained devices exhibit improved thermal performance compared to $SiO_2$/GaN/sapphire HEMT. The lattice temperature was reduced from 507 K in $SiO_2$/GaN/sapphire to 372 K in h-BN/GaN/diamond HEMT. The temperature decrease enhances the drain current and transconductance to 900 mA/mm and 250 mS/mm, corresponding to a 47 % improvement. In addition, the total thermal resistance $R_{th}$ is reduced by a factor of 5 from 27 K.mm/W in GaN/sapphire HEMT to 5.5 K.mm/W in GaN/diamond HEMT. This study indicates that h-BN integration in GaN HEMT as a top heat spreader and a release layer for transfer to diamond substrate can be a promising solution to reduce self-heating effects and extend the device lifetime and reliability.

**Introduction**

In the last few years, GaN HEMTs have gained increasing attention in high-power and high-frequency applications [1,2] thanks to their high breakdown electric field (~10 MV/cm), high electron mobility and excellent saturation velocity ($1.9 \times 10^7$ cm/s) [3–5]. However, self-heating effects induced by the low thermal conductivity of growth substrates, such as silicon and sapphire, limit GaN HEMT performance at high power densities, thus degrading its output power, electron mobility, and reliability. Several studies have been conducted to address self-heating issues, including microfluidic cooling [6], flip-chip bonding [7,8], replacing conventional passivation materials with high thermal conductivity materials (AlN (~ 285 W/mK) [9–11], and $Al_2O_3$ (18-85 W/mK) [12]). In addition to growing GaN HEMTs on high thermal conductivity substrates, for example, SiC (340 W/mK). Yet, self-heating issues have not been managed effectively due to additional challenges raised by the previous solutions, such as lattice mismatch between GaN and growth substrate, reduced thermal conductivity for thicker passivation layers, and design complexity.

Recently, h-BN, a 2D material with a similar structure to graphene, has attracted great interest for its remarkable properties. Its high in-plane thermal conductivity (390 -750 W/mK) [13–16], atomic flatness, and absence of dangling bonds [17] make it a promising heat spreader [18] and a superior dielectric material for GaN HEMTs [19,20]. Furthermore, h-BN has a layered structure, i.e., the atomic layers are bonded through van der Waals interactions, allowing it to be used as a release layer to transfer GaN HEMTs to high thermal conductivity substrates. This method, known as h-BN lift-off, has already proven its success in transferring GaN HEMT to copper and SiC substrates [21,22].

On the other hand, among the solutions reported in the literature to reduce self-heating effects is using diamond as a growth substrate. It is well known that diamond is an ultra-band gap semiconductor having a very high thermal conductivity (~ 2000 W/mK), making it an ideal choice for heat dissipation in GaN HEMT [23–25]. However, the growth of GaN epilayers on diamond substrate using metal-organic chemical vapor deposition (MOCVD) is still challenging because of the thermal coefficient mismatch between the two materials [26,27]. Thus, resulting in wafer cracking during post-epitaxy cooling due to thermal stress accumulation [28]. Therefore, GaN and diamond bonding at room temperature can potentially overcome the previous issues and lead to reduced SHEs in GaN HEMTs [29,30].

In this work, we investigate several approaches to reduce self-heating effects in a GaN HEMT through numerical simulations. First, we simulate and extract the electrical and thermal characteristics of GaN HEMT on sapphire substrate. Second, we study how replacing the $SiO_2$ passivation layer with h-BN can reduce the lattice temperature. Finally, we simulate the electrothermal behavior of an h-BN/GaN HEMT transferred to diamond substrate via the h-BN lift-off technique. The obtained results demonstrate an improvement of 47 % in drain current and transconductance in h-BN/GaN/diamond HEMT and a reduction of thermal resistance by a factor of 5 compared to SiO$_2$/GaN/sapphire HEMT.

## I. Simulation models and structure

Device simulations are performed using Atlas Silvaco software. The GaN HEMT investigated in this work comprises a GaN/AlN/Al$_{0.3}$Ga$_{0.7}$N/GaN structure on a sapphire substrate. The epitaxial layers consist of a 2 μm thick undoped GaN buffer layer. The Al$_{0.3}$Ga$_{0.7}$N barrier layer has a thickness of 20 nm and is sandwiched between a 1 nm AlN thick spacer layer and 2 nm thick cap layer. The devices have a gate length of 1.5 μm, a source-drain spacing of 6 μm, and a gate width of 100 μm. The electron mobility in GaN and AlGaN is 1100 cm$^2$/Vs and 300 cm$^2$/Vs, respectively. The saturation velocity in GaN and AlGaN barrier equals 1.9×10$^7$ and 1.2×10$^7$ cm/s, respectively. Fig. 1a and Fig. 1b show schematic illustrations of SiO$_2$/GaN/sapphire HEMT and h-BN/GaN/sapphire HEMT, respectively. To simulate the substrate effect on heat dissipation in GaN HEMT devices, a thermal electrode is attached to the bottom of the device. The temperature and thermal boundary resistance (TBR) of the electrode are 300 K and 10$^{-7}$- 5×10$^{-7}$ m$^2$K W$^{-1}$ in the case of GaN on sapphire substrate [31,32], and 10$^{-8}$ m$^2$K W$^{-1}$ in the case of GaN on diamond substrate [33,34]. It should be noted that TBR values depend strongly on the interface between the substrate and GaN, which depends, in turn, on the growth method and conditions.

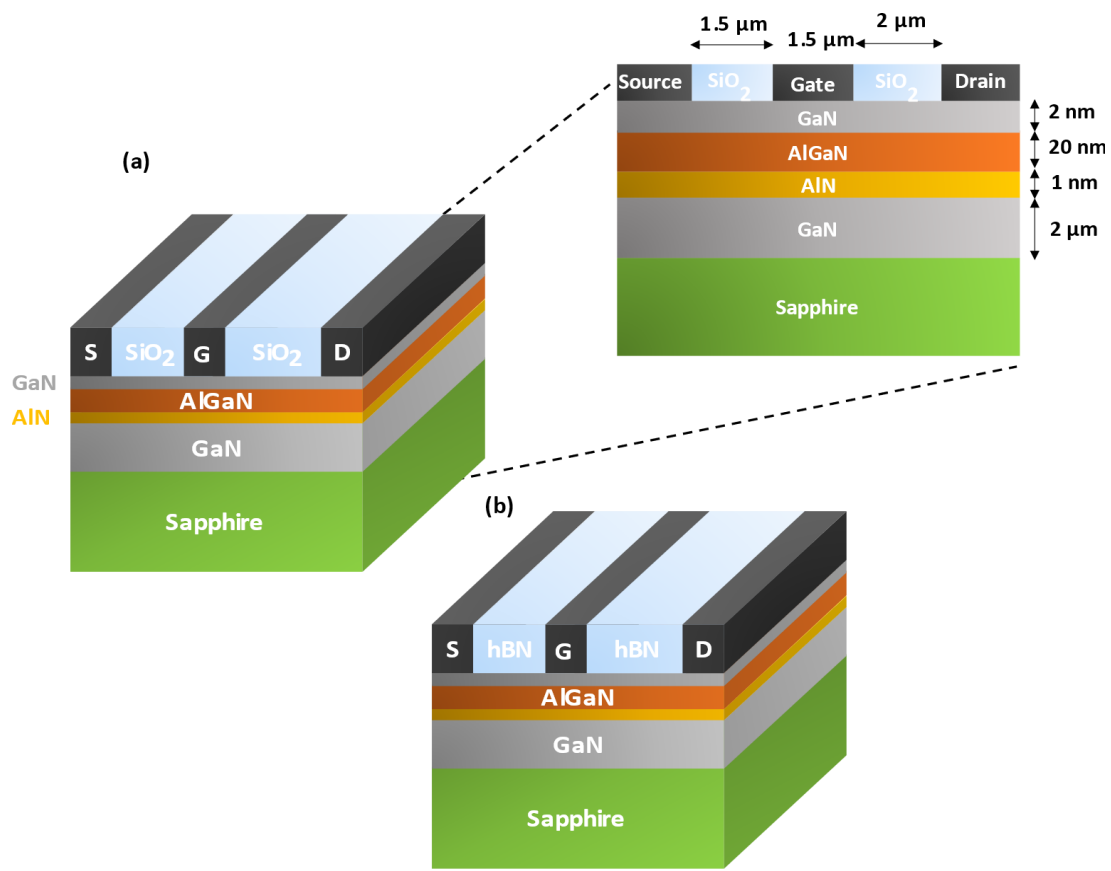

Fig. 1. Schematic view of (a) SiO$_2$/GaN/sapphire HEMT and (b) h-BN/GaN/sapphire HEMT. The onset shows the thickness of each layer and the source-to-drain spacing.

The physical models used in the simulation for mobility, thermal conductivity, heat capacity, recombination, and polarization charges are discussed below:

1. **Low-field mobility model**

In this work, two low-field models are used; the constant low-field mobility described by [35]:

$$\mu_n = \mu_{n0} \left(\frac{T_L}{300}\right)^{-\alpha} \tag{1}$$

Where $\mu_{n0}$ is electron low-field mobility, $T_L$ is the device's actual temperature, and $\alpha$ is a fitting parameter between 1.5-1.8 [35–37].

The second mobility model is the Farahmand Modified Caughey-Thomas model. This model results from fitting the Caughey Thomas model to Monte Carlo data [35,38]:

$$\mu_0(T,N) = \mu_{min}\left(\frac{T_L}{300}\right)^{\beta_1} + \frac{(\mu_{min}-\mu_{max})\left(\frac{T_L}{300}\right)^{\beta_2}}{\left[1+\left[\frac{N}{N_{ref}\left(\frac{T_L}{300}\right)^{\beta_3}}\right]^{\alpha(T/300)^{\beta_4}}\right]} \tag{2}$$

Where N is the total doping density, $\mu_{min}$, $\mu_{max}$, $\alpha$, $\beta_1$, $\beta_1$, $\beta_1$, $\beta_1$ and $N_{ref}$ are parameters extracted from Monte Carlo simulation [35,38]. Al$_{0.3}$Ga$_{0.7}$N mobility parameters are obtained by linear interpolation.

2. **High-field mobility model**

The high-field mobility model enables to predict the material mobility at high electric field, and it is given by [35]:

$$\mu = \frac{\mu_0(T,N) + v^{sat}\frac{E^{n_1-1}}{E_C^{n_1}}}{1+a\left(\frac{E}{E_C}\right)^{n_2}+\left(\frac{E}{E_C}\right)^{n_1}} \tag{3}$$

Where $\mu_0(T,N)$, $E_c$ and $v^{sat}$ are the low field mobility, the critical electric field, and the saturation velocity, respectively. A, $n_1$, and $n_2$ are fitting parameters.

3. **Thermal conductivity model :**

The material thermal conductivity used in the simulation is expressed as [35]:

$$k(T_L) = k(T_0)\left(\frac{T_L}{300}\right)^{-\gamma} \qquad (4)$$

Where $k(T_0)$ is the thermal conductivity at room temperature, $k(T_L)$ is the temperature-dependent thermal conductivity, and γ is a material-dependent parameter. The values of $k(T_0)$ and γ are illustrated in Table 1.

The thermal conductivity and γ of $Al_xGa_{1-x}N$ alloy are computed using the following equations [35]:

$$k_{300}^{AlGaN} = \frac{1}{\frac{x}{k_{300}^{AlN}} + \frac{1-x}{k_{300}^{GaN}} + \frac{x(1-x)}{C}} \qquad (5)$$

$$\gamma^{AlGaN} = x\gamma^{AlN} + (1-x)\gamma^{GaN} \qquad (6)$$

Where x is the content of Al in AlGaN layer, $k_{300}^{GaN}$ and $k_{300}^{AlN}$ are the thermal conductivities of GaN and AlN at room temperature, and C is a bowing factor that accounts for the non-linear variation of thermal conductivity with composition.

### 4. Heat capacity

The dependence of heat capacity on temperature is expressed as [35]:

$$C(T_L) = \rho\left[C_{300} + C_1 \frac{\left(\frac{T_L}{300}\right)^\beta - 1}{\left(\frac{T_L}{300}\right)^\beta + \frac{C_1}{C_{300}}}\right] \qquad (7)$$

Where ρ is the material's mass density, $C_{300}$ is the heat capacity at room temperature, β and $C_1$ are material-dependent parameters.

Table 1 Thermal conductivity and heat capacity parameters used in the simulation.

| Material | Thermal conductivity (W/cm.K) | γ | Density (g/cm³) | Heat capacitance (J/kg.K) |
|---|---|---|---|---|
| GaN | 1.6 [7] | 1.4 [7] | 6.15 [39] | 490 [39] |
| AlN | 1.8 [7] | 1.4 [7] | 3.23 [35] | 600 [35] |
| AlGaN | Eq. (5) | Eq. (6) | Linear interpolation | Linear interpolation |
| Sapphire | 0.49 [7] | 1 [7] | 3.98 [39] | 770 [39] |
| Diamond | 11.48 (in-plane) [40] | 0.55 [40] | 3.51 [41] | 520 [41] |
| h-BN | 4 (in-plane) [13] | - | 2.28 [13] | 810 [42] |

## 5. Polarization charge density

The polarization charge density is calculated using the equations reported in [43,44]:

$$P_{sp}^{AlGaN} = xP_{sp}^{AlN} + (1-x)P_{sp}^{GaN} + 0.038x(1-x) \qquad (8)$$

$$P_{pz}^{GaN} = -0.0918\varepsilon + 9.541\varepsilon^2$$

$$P_{pz}^{AlN} = -1.808\varepsilon + 5.624\varepsilon^2 \text{ for } \varepsilon < 0$$

$$P_{pz}^{AlN} = -1.808\varepsilon - 7.888\varepsilon^2 \text{ for } \varepsilon >$$

$$P_{pz}^{AlGaN} = xP_{pz}^{AlN} + (1-x)P_{pz}^{GaN}$$

Where $\varepsilon$ is the basal strain induced in the AlGaN layer and is equal to $\varepsilon = \frac{a_s - a(x)}{a(x)}$, with $a_s$ and $a(x)$ are the lattice constants of the substrate and the unstrained AlGaN alloy. The polarization charge density is distributed at GaN/AlN, AlN/AlGaN, and AlGaN/GaN interfaces over a thickness of 0.5 nm, as reported in [45].

## 6. Shockley-Read Hall recombination model

The dominant recombination in III-nitride materials is Shockley-Read Hall (SRH) recombination, which describes the recombination of electrons and holes due to defects present in the material band gap and is given by [35]:

$$R_{RSH} = \frac{np - n_i^2}{\tau_p \left[n + n_i \exp\left(\frac{E_{Fi} - E_T}{kT_L}\right)\right] + \tau_n \left[p + n_i \exp\left(\frac{-(E_{Fi} - E_T)}{kT_L}\right)\right]}$$

Where $E_T$ is the trap energy level within the band gap, $E_{Fi}$ is the intrinsic Fermi level, $n_i$ is the intrinsic carrier density and $\tau_n$, $\tau_p$ are the electron and hole lifetimes equal to 5 ns.

## I. Results and discussion

First, we simulate the transfer and DC characteristics of GaN HEMT with and without the self-heating effect. The gate voltage $V_{GS}$ is swept from -8 to 0 V and $V_{DS}$ from 0 to 15 V. Fig. 2a and Fig. 2b show the transfer characteristic of SiO$_2$/GaN/sapphire HEMT with and without the SHE at $V_{DS}$ = 8 V. The drain current at $V_{GS}$ = 0 V reaches 1056 mA/mm without SHE and decreases to 683 mA/mm when considering the SHE effect. In addition, the maximum transconductance also shows a reduction of 36.65 % from 281 mS/mm to 178 mS/mm, as illustrated in Fig. 2b.

Furthermore, the DC characteristics of SiO$_2$/GaN/sapphire without and with SHE are shown in Fig. 2c and Fig. 2d, respectively. Because of SHE induced by the low thermal conductivity of sapphire substrate, the drain current decreases as the applied voltage increases. The maximum drain current, $I_{DS, max}$ decreases from 1100 to 598 mA/mm at $V_{DS}$ = 15 V and $V_{GS}$ = 0 V, which is about a 45.5 % reduction.

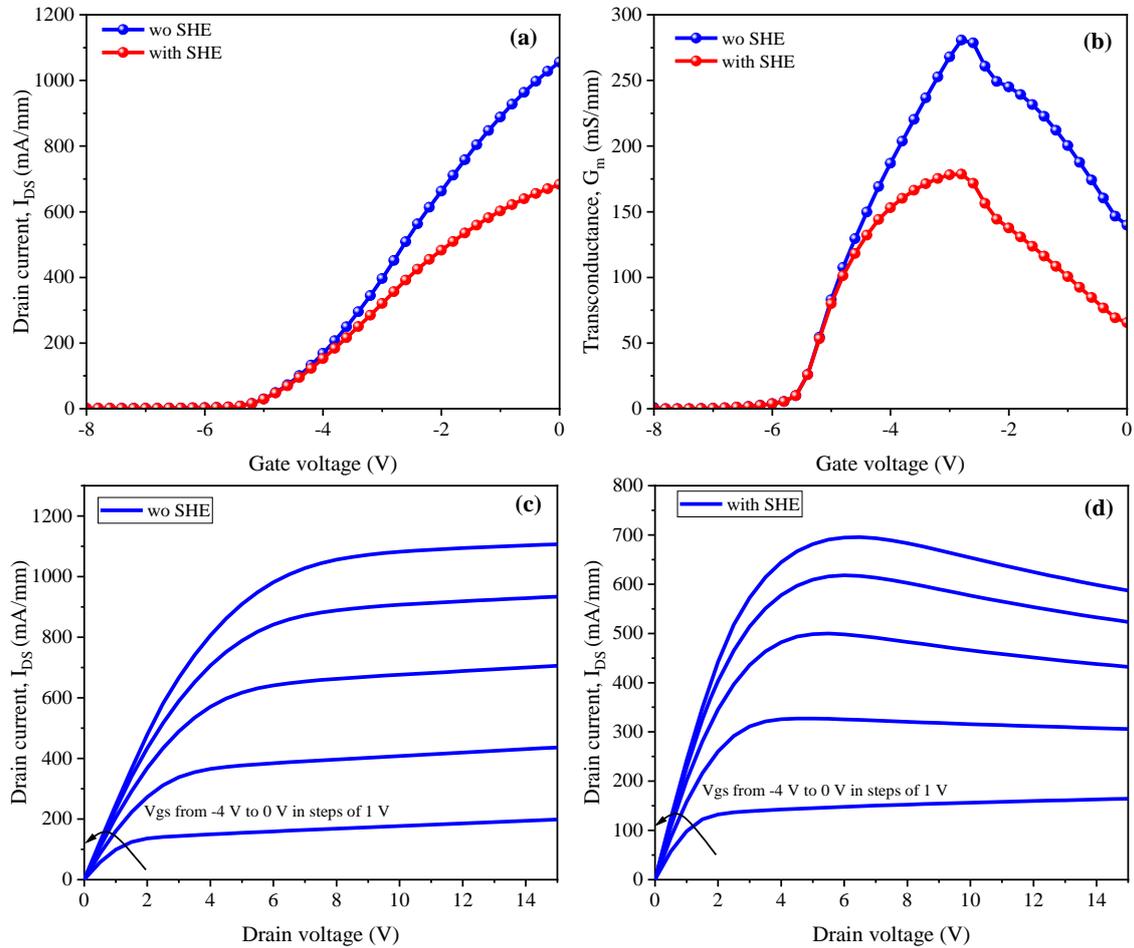

Fig. 2. (a) I-V characteristic of SiO$_2$/GaN/sapphire HEMT with and without (wo) SHE at $V_{DS}$ = 8 V. (b) The corresponding transconductance with and without (wo) SHE. (c) DC characteristic of SiO$_2$/GaN/sapphire HEMT without SHE. (d) DC characteristic of SiO$_2$/GaN/sapphire HEMT with SHE.

This decrease in current and transconductance is due mainly to an increase in the lattice temperature. Fig. 3a and Fig. 3b show lateral and vertical temperatures of GaN HEMT on sapphire substrate with SiO$_2$ as a passivation layer. As can be seen, the SHE results in a high temperature, reaching a maximum value of 507 K at the gate edge near the drain side. The elevated lattice temperature causes an increase in phonon scattering, degrading the electron mobility and device performance.

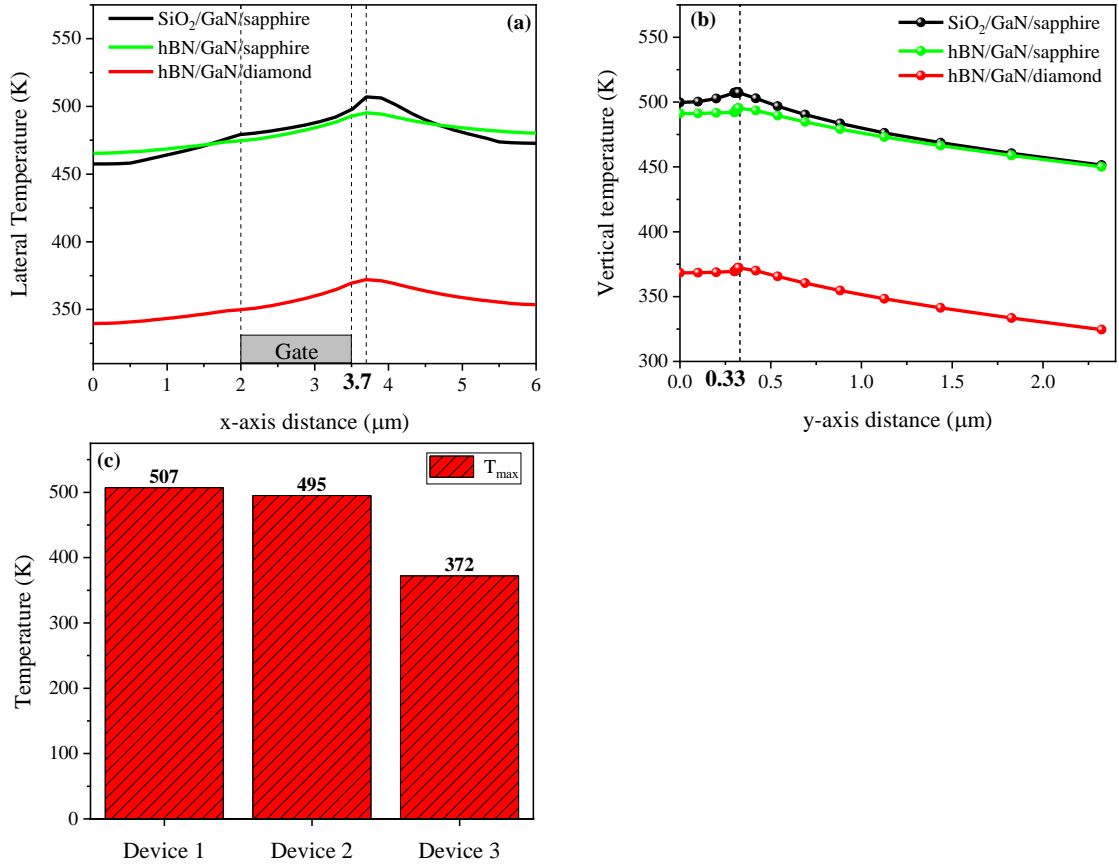

Fig. 3. (a) Lateral and (b) vertical temperatures of SiO$_2$/GaN/sapphire (Device 1), h-BN/GaN/sapphire (Device 2) and h-BN/GaN/diamond HEMT (Device 3). (c) Maximum lattice temperature for each of the three devices.

To reduce self-heating effects caused by the low thermal conductivity of sapphire substrate, first, we have replaced the SiO$_2$ passivation layer with h-BN. We believe that h-BN can act as an efficient top heat spreader in GaN HEMTs owing to its high lateral thermal conductivity compared to SiO$_2$ and Si$_3$N$_4$. In addition, h-BN can be employed as a gate dielectric for GaN HEMTs, as reported by previous studies in the literature [19,20]. The lateral and vertical temperatures of h-BN/GaN/sapphire HEMT and its temperature gradient profile are illustrated in Fig. 3a, Fig. 3b, and Fig. 4b, respectively. The lattice temperature was reduced from 507 K to 495 K after replacing SiO$_2$ with h-BN thanks to the high thermal conductivity of h-BN compared to SiO$_2$.

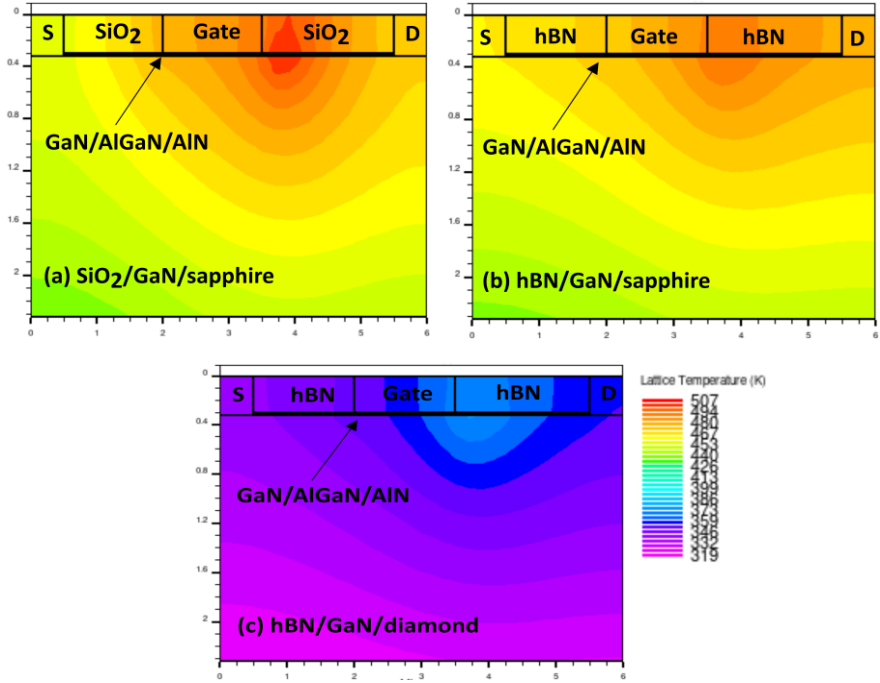

Fig. 4 (a) Temperature gradient profile of SiO$_2$/GaN/sapphire HEMT, (b) h-BN/GaN/sapphire HEMT, (c) h-BN/GaN/diamond HEMT.

Second, for additional reduction of the lattice temperature and improvement of GaN HEMT performance, we propose in this study to transfer GaN HEMT from sapphire substrate to diamond substrate via lift-off transfer technique. This method allows an easy transfer and bonding of GaN epitaxial layers to any other substrate. Furthermore, the bonding of GaN to diamond substrate is achieved at room temperature, enabling us to avoid the issues induced during the growth of GaN epilayers at high temperatures using MOCVD technique, such as wafer cracking, delamination, and void formation across the GaN/diamond interface.

The idea consists of growing GaN epilayers on h-BN/sapphire template. The device is then processed using photolithography, inductively coupled plasma (ICP) dry etching, chemical etching, and e-beam evaporation to deposit the gate, source, and drain electrodes. A commercial water-dissolvable tape is placed on the top of the device. By applying a mechanical force on the tape, the device is released from sapphire substrate and is bonded to diamond substrate (see Fig. 5). As it is known, the roughness of diamond is high compared to other substrates, making the bonding at room temperature challenging. To address this issue, a benzo-cyclobutene (BCB) polymer can be used as an interfacial layer to enhance the adhesion of GaN to diamond substrate, as proposed in a previous study [22].

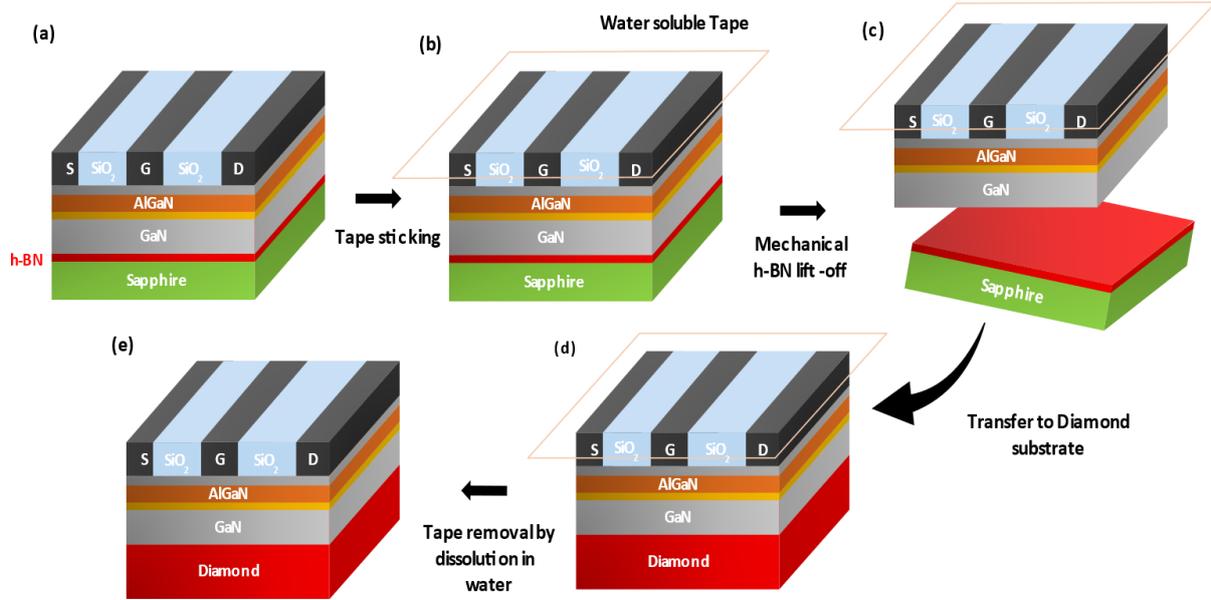

Fig. 5. Schematic illustration of GaN HEMT transfer using h-BN lift-off. (a) GaN HEMT structure on h-BN/sapphire template grown by MOCVD. (b) GaN HEMT with water-dissolvable tape on its top. (c) h-BN lift-off of GaN HEMT from sapphire substrate. (d) GaN HEMT bonding to diamond substrate. (e) Tape dissolution in warm water.

To simulate the thermal behavior of diamond substrate, we have placed a thermal electrode at the GaN/diamond interface with a thermal boundary resistance of $10^{-8}$ m$^2$K W$^{-1}$. This value is consistent with reported results of TBR measured for GaN layers bonded to diamond at room temperatures using different transfer techniques. The transfer and DC characteristics of GaN HEMT transferred to diamond substrate with h-BN as passivation layer are illustrated in Fig. 6. After transfer, the drain current increases to 900 mA/mm at $V_{GS}$ = 0 V compared to 683 mA/mm in the case of SiO$_2$/GaN/sapphire HEMT, corresponding to 47 % improvement. The transconductance of h-BN/GaN/diamond HEMT also rises to 250 mS/mm compared to only 180 mS/mm. As for the DC characteristic of GaN/diamond HEMT, the drain current collapse at high bias voltage is significantly lowered thanks to the lattice temperature reduction. The maximum lattice temperature is decreased from 507 to 372 K, as shown in Fig. 3 and Fig. 4. These results are attributed to the improved heat dissipation caused by the high thermal conductivity of diamond and the lift-off transfer process. This latter allows a good adhesion between GaN epilayers and substrate and a smooth interface between the two.

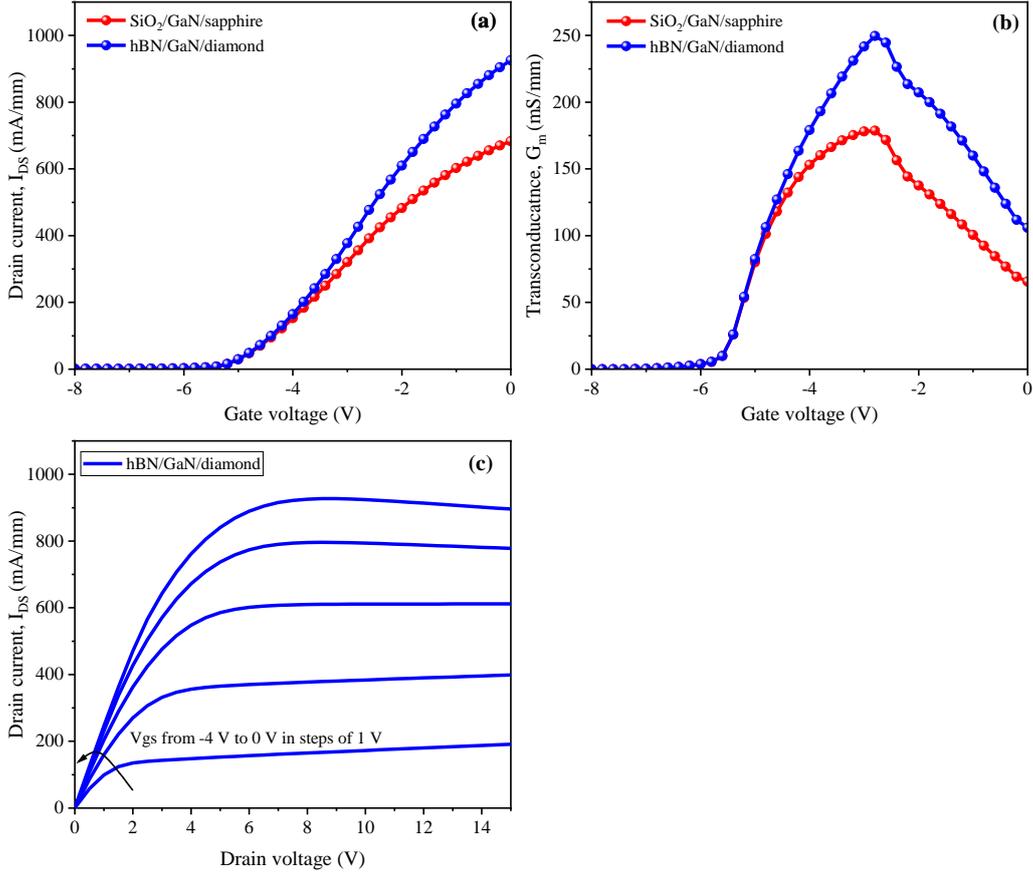

Fig. 6 (a) I-V characteristics of GaN HEMT on sapphire and diamond with h-BN layer at $V_{DS}$ = 8 V. (b) The corresponding transconductance h-BN/GaN/sapphire and h-BN/GaN/diamond HEMTs at $V_{DS}$ = 8 V. (c) DC characteristics of h-BN/GaN/diamond HEMT.

To better visualize the effect of temperature on GaN HEMT performance, we have simulated the DC characteristics at high bias voltage ranging from 0 to 40 V (see Fig. 7a). Fig. 7b shows the drain current reduction versus drain voltage in the case of $SiO_2$/GaN/sapphire and h-BN/GaN/diamond HEMTs. The current reduction is given by $I_{DS,reduction}(\%) = \frac{I_{DS,max} - I_{DS-VDS}}{I_{DS,max}} \times 100$, where $I_{DS,max}$ is the maximum drain current achieved at $V_{DS}$ = 10 V and $I_{DS-VDS}$ is the drain current at a specific $V_{DS}$ where we want to calculate the current drop. At $V_{DS}$ = 40 V, the drain current drop is 18 % in h-BN/GaN/diamond HEMT, while it reaches 46 % in $SiO_2$/GaN/sapphire HEMT. This demonstrates that GaN on diamond HEMT can operate at high power densities while maintaining a low channel temperature. Contrary to GaN on sapphire HEMT, whose drain current significantly decreases to half its initial value at $V_{DS}$ = 40 V.

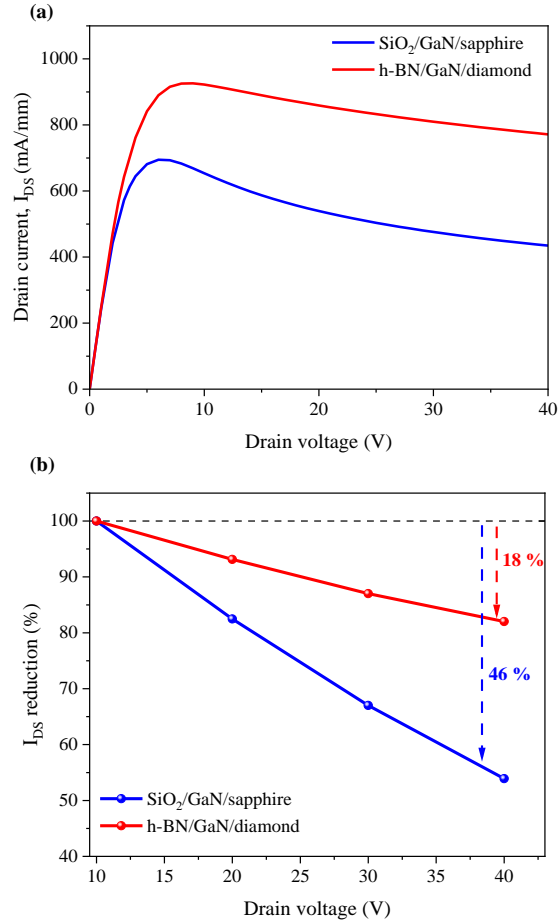

Fig. 7. (a) The drain current of h-BN/GaN/diamond (red line) and SiO$_2$/GaN/sapphire (blue line) HEMTs. (b) The drain current reduction in h-BN/GaN/diamond (red line) and h-BN/GaN/diamond HEMTs (blue line) at $V_{GS}$ = 0 V.

Additionally, Fig. 8 shows the variation of lattice temperature as a function of the dissipated power density at $V_{GS}$ = 0 V. From the figure, it is clear that temperature rise is rapid in SiO$_2$/GaN/sapphire HEMT where it reaches 771 K at $P_{diss}$ = 17.4 W/mm compared to only 390 K in h-BN/GaN/diamond HEMT. The total thermal resistance $\left(R_{th} = \frac{\Delta T_{lattice}}{\Delta P_{diss}}\right)$ was calculated from the slope of temperature variation versus power density for GaN/sapphire (27 K.mm/W) and GaN/diamond (5.5 K.mm/W) HEMTs. The obtained thermal resistances closely match the values reported for GaN/diamond (6.7 K.mm/W, 6.5 K.mm/W) [23,46] and for GaN/sapphire HEMT (32.9 K.mm/W) [47]. The total thermal resistance is 5 times less in h-BN/GaN/diamond than in SiO$_2$/GaN/sapphire HEMT. The thermal resistance values achieved in this study are lower than those reported in the literature, whether for GaN/sapphire or GaN/diamond HEMTs. This can be due to the large area of these devices, which leads to improved heat dissipation and, thus, lower thermal resistance.

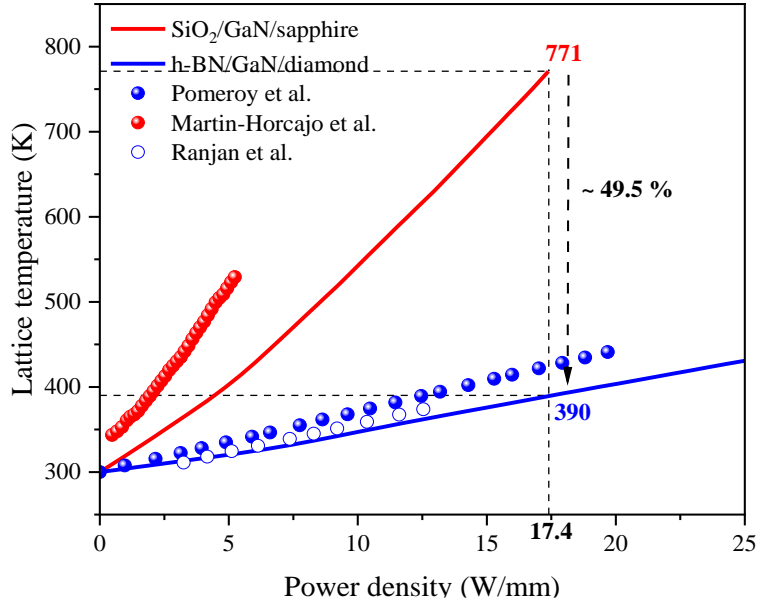

Fig. 8 Lattice temperature variation versus power density at $V_{GS}$ = 0 V. Solid lines represent data obtained using simulation, while dots are data reported by previous studies. Red dots denote the lattice temperature variation for GaN HEMT on sapphire reported by Martin-Horcajo et al. [47]. Blue dots are attributed to GaN HEMT on diamond lattice temperature achieved by Rajan et al. [23] (empty dots) and Pomeroy et al. [46] (filled dots).

To evaluate the dynamic performance of $SiO_2$/GaN/sapphire and h-BN/GaN/diamond HEMTs, we have performed transient simulations for the drain current and temperature at $V_{GS}$ = 0 V and $V_{DS}$ switching from 0 to 15 V (see). Fig. 9 shows the transient drain current and lattice temperature of $SiO_2$/GaN/sapphire and h-BN/GaN/diamond HEMTs. As the device is switched ON, in GaN on sapphire HEMT, the drain current increases rapidly, reaching 1350 mA/mm, then decreases to 587 mA/mm, about 56 % reduction. Whereas, in GaN on diamond HEMT, the current decreases to only 918 mA/mm, corresponding to 32 % decrease (see Fig. 9a and Fig. 9b). The lattice temperature increases rapidly from 300 to 476 K during 800 ns for GaN/sapphire HEMT, as shown in Fig. 9c. This is due to the drastic increase of current which induces a large amount of power dissipation and heat generation. Furthermore, a knee point is observed at 1 µs, where the temperature reaches 500 K. The rising time for GaN on sapphire HEMT is about 1.8 µs and the maximum temperature is about 507 K, which remains constant despite increasing the simulation time. After switching the device OFF, the temperature decreases quickly to 323 K (1.7 µs). Then, a knee point is noticed from 21.7 to 23.3 µs where the temperature becomes 300 K. The falling time is estimated to be 3.3 µs.

The same scenario occurs for GaN on diamond HEMT, i.e., the temperature increases rapidly from 300 to 363 K during 290 ns. Then, a knee point is noticed at 460 ns, where the temperature rises to 367 K. The rising time is about 750 ns, and the maximum temperature is 367 K. As the device is switched OFF, the temperature falls abruptly to 304 K. This is followed by a knee point at 430 ns, where the temperature reaches 300 K. The falling time is about 790 ns. As can be noticed, the rising and falling time in the case of GaN on sapphire HEMT is higher than GaN on diamond HEMT. This indicates that the former cannot regain its initial state due to increased thermal resistance, reduced carrier mobility, and current density caused by SHEs.

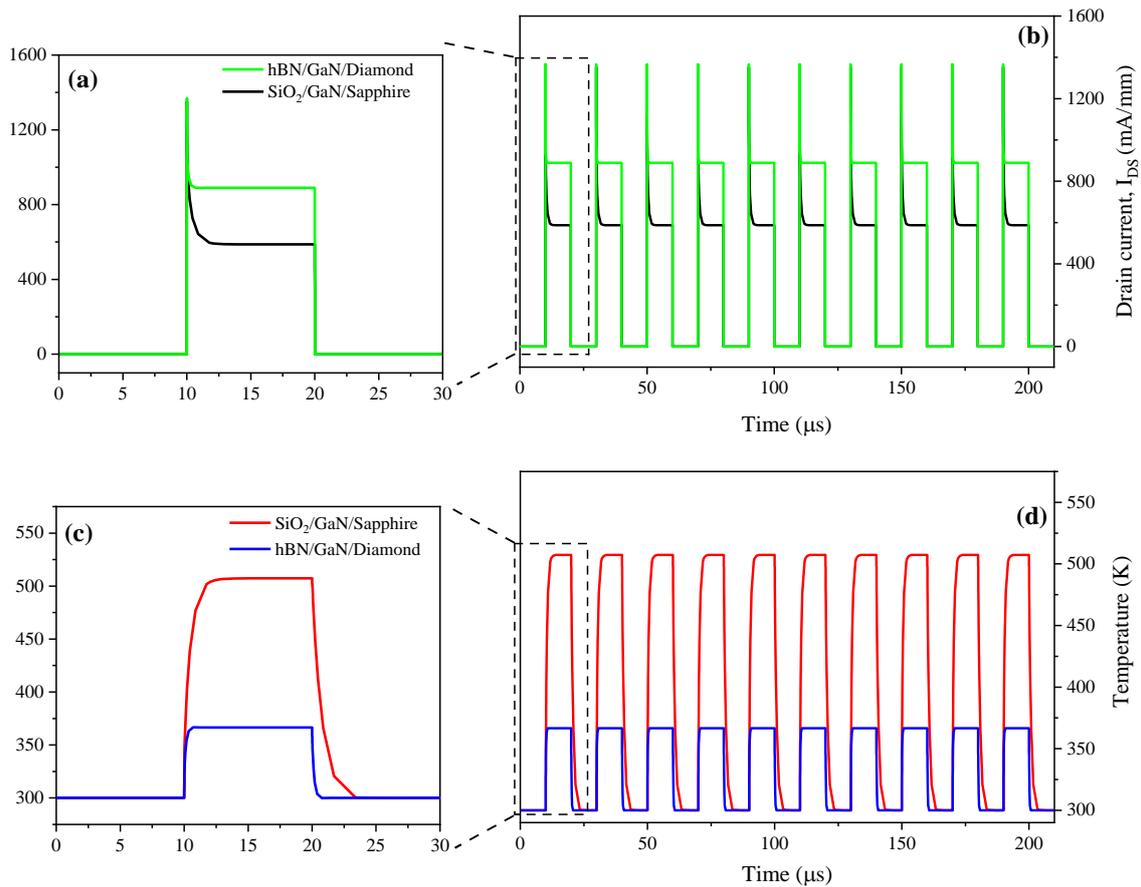

Fig. 9 Transient simulation of drain current (a, b) and temperature (c, d) at $V_{GS} = 0$ V and $V_{DS}$ switching from 0 to 15 V.

**Conclusion**

In this paper, we investigate through numerical simulations the passivation of GaN HEMT with h-BN instead of $SiO_2$ and its transfer from sapphire substrate to diamond to reduce SHEs. The obtained results show that replacing $SiO_2$ passivation layer with h-BN enables reducing the junction temperature from 507 to 495 K. Furthermore, h-BN/GaN/diamond HEMT exhibited a low junction temperature of about 372 K compared to 507 K in the case of $SiO_2$/GaN/sapphire HEMT. The drain current reduction at a bias voltage was estimated to be 18 % in GaN/diamond, while it was 46 % in $SiO_2$/GaN/sapphire HEMT. In addition, the extracted thermal resistance in GaN/sapphire (27 K.mm/W) was 5 times higher than in GaN/diamond HEMT (5.5 K.mm/W). Furthermore, transient simulations showed that h-BN/GaN/diamond HEMT exhibits a reduced falling time (750 ns) compared to $SiO_2$/GaN/sapphire HEMT (3.3 µs). These findings indicate that h-BN/GaN/diamond HEMT can be a promising device for high-frequency switching applications such as DC-DC and DC-AC converters.

**Declaration of competing interest**

The authors declare that they have no competing financial interests or personal relationships that could have appeared to influence the work reported in this paper.

**Acknowledgments**

We acknowledge funding from the French CNRS IRN ATLAS.

**References**

[1] T. Palacios, A. Chakraborty, S. Rajan, C. Poblenz, S. Keller, S.P. DenBaars, J.S. Speck, U.K. Mishra, High-power AlGaN/GaN HEMTs for Ka-band applications, IEEE Electron Device Lett. 26 (2005) 781–783.

[2] R. C. Fitch; D. E. Walker; A. J. Green; S. E. Tetlak; J. K. Gillespie, R. D. Gilbert, K. A. Sutherlin, W. D. Gouty, J. P. Theimer, G. D. Via, K. D. Chabak, G. H. Jessen, Implementation of High-Power-Density X-Band AlGaN/GaN High Electron Mobility Transistors in a Millimeter-Wave Monolithic Microwave Integrated Circuit Process, IEEE Electron Device Lett. 36 (2015) 1004–1007.

[3] U. K. Mishra, P. Parikh, Y. F. Wu, AlGaN/GaN HEMTs - An overview of device operation and applications, Proc. IEEE 90 (2002) 1022–1031.

[4] U. K. Mishra, L. Shen, T. E. Kazior, Y. F. Wu, GaN-based RF power devices and amplifiers, Proc. IEEE 96 (2008) 287–305.

[5] H. Amano Y. Baines, E. Beam, M. Borga, T. Bouchet, P. R. Chalker, M. Charles, K. J. Chen, N. Chowdhury, R. Chu, C. de Santi, M. M. De Souza, S. Decoutere, L. di Cioccio, B. Eckardt, T. Egawa, P. Fay, J. J. Freedsman, L. Guido, O. Häberlen, G. Haynes, T. Heckel, D. Hemakumara, P. Houston, J. Hu, M. Hua, Q. Huang, A. Huang, S. Jiang, H. Kawai, D. Kinzer, M. Kuball, A. Kumar, K. B. Lee, X. Li, D. Marcon, M. März, R. McCarthy, G. Meneghesso, M. Meneghini, E. Morvan, A. Nakajima, E. M. S. Narayanan, S. Oliver, Tomás Palacios, D. Piedra, M. Plissonnier, R. Reddy, M. Sun, I. Thayne, A. Torres, N. Trivellin, V. Unni, M. J. Uren, M. V. Hove, D. J. Wallis, J. Wang, J. Xie, S. Yagi, S. Yang, C. Youtsey, R. Yu, E. Zanoni, S. Zeltner, Y. Zhang, The 2018 GaN power electronics roadmap, J. Phys. D Appl. Phys. 51 (2018) 163001.

[6] Y. Ye, M. Wu, Y. Kong, R. Liu, L. Yang, X. Zheng, B. Jiao, X. Ma, W. Bao, Y. Hao, Active Thermal Management of GaN-on-SiC HEMT With Embedded Microfluidic Cooling, IEEE Trans. Electron Devices 69 (2022) 5470–5475.

[7] J. Das, H. Oprins, H. Ji, A. Sarua, W. Ruythooren, J. Derluyn, M. Kuball, M. Germain, G. Borghs, Improved thermal performance of AlGaN/GaN HEMTs by an optimized flip-chip design, IEEE Trans. Electron Device 53 (2006) 2696–2702.

[8] J. Sun, H. Fatima, A. Koudymov, A. Chitnis, X. Hu, H.M. Wang, J. Zhang, G. Simin, J. Yang, M.A. Khan, Thermal management of AlGaN-GaN HFETs on sapphire using flip-chip bonding with epoxy underfill, IEEE Electron Device Lett. 24 (2003) 375–377.

[9] G. A. Slack, R. A. Tanzilli, R. O. Pohl, J. W. Vandersande, The intrinsic thermal conductivity of AlN, J. Phys. Chem. Solids 48 (1987) 641–647.

[10] N. Tsurumi, H. Ueno, T. Murata, H. Ishida, Y. Uemoto, T. Ueda, K. Inoue, T. Tanaka, AlN passivation over AlGaN/GaN HFETs for surface heat spreading, IEEE Trans. Electron Device 57 (2010) 980–985.

[11] P. Murugapandiyan, D. Nirmal, M. T. Hasan, A. Varghese, J. Ajayan, A. S. Augustine Fletcher, N. Ramkumar, Influence of AlN passivation on thermal performance of AlGaN/GaN high-electron mobility transistors on sapphire substrate: A simulation study, Mat. Sci. and Engineering: B 273 (2021) 115449.

[12] N. Bresson, S. Cristoloveanu, C. Mazuré, F. Letertre, H. Iwai, Integration of buried insulators with high thermal conductivity in SOI MOSFETs: Thermal properties and short channel effects, Solid State Electron. 49 (2005) 1522–1528.

[13] E. K. Sichel, R. E. Miller, M. S. Abrahams, C. J. Buiocchi, Heat capacity and thermal conductivity of hexagonal pyrolytic boron nitride, Phys. Rev. B 13 (1976) 4607–4611.

[14] Q. Cai, D. Scullion, W. Gan, A. Falin, S. Zhang, K. Watanabe, T. Taniguchi, Y. Chen, E. J. G. Santos, L. H. Li, High thermal conductivity of high-quality monolayer boron nitride and its thermal expansion. Sci. adv. 5 (2019), eaav0129.

[15] I. Jo, M. T. Pettes, J. Kim, K. Watanabe, T. Taniguchi, Z. Yao, L. Shi, Thermal conductivity and phonon transport in suspended few-layer hexagonal boron nitride, Nano Lett. 13 (2013), 550-554.

[16] C. Wang, J. Guo, L. Dong, A. Aiyiti, X. Xu, B. Li, Superior thermal conductivity in suspended bilayer hexagonal boron nitride, Sci. Rep. 6 (2016), 25334.

[17] D. Choi, N. Poudel, S. Park, D. Akinwande, S. B. Cronin, K. Watanabe, T. Taniguchi, Z. Yao, L. Shi, Large Reduction of Hot Spot Temperature in Graphene Electronic Devices with Heat-Spreading Hexagonal Boron Nitride, ACS Appl. Mater. Interfaces 10 (2018) 11101–11107.

[18] D. Jeon, J. Lim, J. Bae, A. Kadirov, Y. Choi, S. Lee, Suppression of self-heating in nanoscale interfaces using h-BN based anisotropic heat diffuser, Appl. Surf. Sci. 543 (2021) 148801.

[19] J. C. Gerbedoen, A. Soltani, M. Mattalah, M. Moreau, P. Thevenin, J. C. De Jaeger, AlGaN/GaN MISHEMT with hBN as gate dielectric. Diam. Relat. Mater. 18 (2009) 1039–1042.

[20] B. Ren, M. Liao, M. Sumiya, J. Li, L. Wang, X. Liu, Y. Koide, L. Sang, Layered boron nitride enabling high-performance AlGaN/GaN high electron mobility transistor, J. Alloys Compd. 829 (2020) 154542.

[21] M. Hiroki, K. Kumakura, Y. Kobayashi, T. Akasaka, T. Makimoto, H. Yamamoto, Suppression of self-heating effect in AlGaN/GaN high electron mobility transistors by substrate-transfer technology using h-BN, Appl. Phys. Lett. 105 (2014) 193509.

[22] M. J. Motala, E. W. Blanton, A. Hilton, E. Heller, C. Muratore, K. Burzynski, J. L. Brown, K. Chabak, M. Durstock, M. Snure, N. R. Glavin, Transferrable AlGaN/GaN High-Electron Mobility Transistors to Arbitrary Substrates via a Two-Dimensional Boron Nitride Release Layer, ACS Appl. Mater. Interfaces 12 (2020) 21837–21844.

[23] K. Ranjan, S. Arulkumaran, G. I. Ng, A. Sandupatla, Investigation of Self-Heating Effect on DC and RF Performances in AlGaN/GaN HEMTs on CVD-Diamond, IEEE J. Electron Device Society 7 (2019) 1264–1269.

[24] M. J. Tadjer, T. J. Anderson, M. G. Ancona, P. E. Raad, P. Komarov, T. Bai, J. C. Gallagher, A. D. Koehler, M. S. Goorsky, D. A. Francis, K. D. Hobart, F. J. Kub, GaN-On-Diamond HEMT Technology with $T_{AVG}$ = 176 °C at $P_{DC, max}$ = 56 W/mm Measured by Transient Thermoreflectance Imaging, IEEE Electron Device Lett. 40 (2019) 881–884.

[25] T. Zhu, X. F. Zheng, Y. R. Cao, C. Wang, W. Mao, Y. Z. Wang, M. H. Mi, M. Wu, J. H. Mo, X. H. Ma, Y. Hao, Study on the effect of diamond layer on the performance of double-channel AlGaN/GaN HEMTs, Semicond. Sci. Technol. 35 (2020) 055006.

[26] R. R. Reeber, K. Wang, Lattice parameters and thermal expansion of GaN, J. Mater. Research 15 (2000) 40–44.

[27] C. Moelle, S. Klose, F. Szücs, H. J. Fecht, C. Johnston, P. R. Chalker, M. Werner, Measurement and calculation of the thermal expansion coefficient of diamond, Diam. Relat. Mater. 6 (1997) 839–842.

[28] S. Choi, E. Heller, D. Dorsey, R. Vetury, S. Graham, The impact of mechanical stress on the degradation of AlGaN/GaN high electron mobility transistors, J. Appl. Phys. 114 (2013) 164501.

[29] M. D. Smith, J. A. Cuenca, D. E. Field, Y. C. Fu, C. Yuan, F. Massabuau, S Mandal, J. W. Pomeroy, R. A. Oliver, M. J. Uren, K. Elgaid, O. A. Williams, I. Thayne, M. Kuball, GaN-on-diamond technology platform: Bonding-free membrane manufacturing process, AIP Adv. 10 (2020) 035306.

[30] Z. Cheng, F. Mu, L. Yates, T. Suga, S. Graham, Interfacial Thermal Conductance across Room-Temperature-Bonded GaN/Diamond Interfaces for GaN-on-Diamond Devices, ACS Appl. Mater. Interfaces 12 (2020) 8376–8384.

[31] A. Sarua, H. Ji, K.P. Hilton, D.J. Wallis, M.J. Uren, T. Martin, M. Kuball, Thermal boundary resistance between GaN and substrate in AlGaN/GaN electronic devices, IEEE Trans. Electron Devices. 54 (2007) 3152–3158.

[32] V. O. Turin. A. A. Balandin, Performance degradation of GaN field-effect transistors due to thermal boundary resistance at GaN/substrate interface, Electronics Lett. 40 (2004) 81–83.

[33] Z. Cheng, F. Mu, L. Yates, T. Suga, S. Graham, Interfacial Thermal Conductance across Room-Temperature-Bonded GaN/Diamond Interfaces for GaN-on-Diamond Devices, ACS Appl. Mater. Interfaces 12 (2020) 8376–8384.

[34] F. Z. Tijent, M. Faqir, H. Chouiyakh, E. H. Essadiqi, Review-Integration Methods of GaN and Diamond for Thermal Management Optimization, ECS J. Solid State Sci. Technol 10 (2021) ac12b3.

[35] Atlas Silvaco, User's Manual Device Simulation Software, Silvaco Inc. Santa Clara, CA, USA, 2016.

[36] X. Z. Dang, P. M. Asbeck, E. T. Yu, G. J. Sullivan, M. Y. Chen, B. T. McDermott, K. S. Boutros, J. M. Redwing, Measurement of drift mobility in AlGaN/GaN heterostructure field-effect transistor, Appl. Phys. Lett. 74 (1999) 3890–3892.

[37] F. Z. Tijent, M. Faqir, P. L. Voss, H. Chouiyakh, E. H. Essadiqi, An analytical model to calculate the current–voltage characteristics of AlGaN/GaN HEMTs, J. Comput. Electron 21 (2022) 644–653.

[38] M. Farahmand, C. Garetto, E. Bellotti, K. F. Brennan, M. Goano, E. Ghillino, G. Ghione, J. D. Albrecht, P. P. Ruden, Monte Carlo simulation of electron transport in the III-nitride Wurtzite phase materials system: Binaries and ternaries, IEEE Trans. Electron Device 48 (2001) 535–542.

[39] M. Kuball, G. J. Riedel, J. W. Pomeroy, A. Sarua, M. J. Uren, T. Martin, K. P. Hilton, J. O. Maclean, D. J. Wallis, Time-resolved temperature measurement of AlGaN/GaN electronic devices using micro-raman spectroscopy, IEEE Trans. Electron Device 28 (2007) 86–89.

[40] H. Guo, Y. Kong, T. Chen, Thermal simulation of high power GaN-on-diamond substrates for HEMT applications, Diam. Relat. Mater. 73 (2017) 260–266.

[41] A. Wang, M. J. Tadjer, F. Calle, Simulation of thermal management in AlGaN/GaN HEMTs with integrated diamond heat spreaders, Semicond. Sci. Technol. 28 (2013) 055010.

[42] A. S. Dworkin, D. J. Sasmor, E. R. Van Artsdalen, The Thermodynamics of Boron Nitride; Low-Temperature Heat Capacity and Entropy; Heats of Combustion and Formation, J. Chem. Phys. 22 (1954) 837–842.

[43] O. Ambacher, J. Majewski, C. Miskys, A. Link, M. Hermann, M. Eickhoff, M. Stutzmann, F. Bernardini, V. Fiorentini, V. Tilak, B. Schaff, L. F. Eastman, Pyroelectric properties of Al(In)GaN/GaN hetero- and quantum well structures, J. Phys. Condens. Matter 14 (2002) 3399–3434.

[44] V. Fiorentini, F. Bernardini, O. Ambacher, Evidence for non-linear macroscopic polarization in III-V nitride alloy heterostructures, Appl. Phys. Lett. 80 (2002) 1204–1206.

[45] M. Faqir, G. Verzellesi, G. Meneghesso, E. Zanoni, F. Fantini, Investigation of high-electric-field degradation effects in AlGaN/GaN HEMTs, IEEE Trans. Electron Device 55 (2008) 1592–1602.

[46] J. Pomeroy, M. Bernardoni, A. Sarua, A. Manoi, D. C. Dumka, D. M. Fanning, M. Kuball, Achieving the best thermal performance for GaN-on-diamond. Technical Digest – IEEE Compound Semiconductor Integrated Circuit Symposium, CSIC. (2013).

[47] S. M. -Horcajo, A. Wang, M. F. Romero, M. J. Tadjer, F. Calle, Simple and accurate method to estimate channel temperature and thermal resistance in AlGaN/GaN HEMTs, IEEE Trans. Electron Devices 60 (2013) 4105–4111.